\documentclass[aps,
groupedaddress,showpacs,nofootinbib,prl,twocolumn]{revtex4}
\usepackage{amssymb,amsmath}
\usepackage{graphics}
\usepackage[english]{babel}
\usepackage{graphicx}
\usepackage{dcolumn}
\usepackage{bm}

\def\ins#1{{\it #1}}
\def\sdag{\dagger}

\def\meq#1{}
\def\meq#1{}

\def\cc{{\rm c.c.}}

\def\ins#1{}

\def\comment#1{}
\newcommand{\lfrac}[2]{#1/#2}

\def\mn#1{\marginpar[]{\scriptsize#1}}
\def\mn#1{}

\def \xp{x'}

\def\nablab{\mbox{\boldmath$\nabla$}}

\usepackage{ulem}
\usepackage{cancel}
\usepackage{color}
\def\rr#1{\textcolor{red}{#1}}

\def\mn#1{\marginpar[\tiny{\rr{#1}}]{\tiny{\rr{#1}}}}

\def\rr#1{}

\usepackage{hyperref}

\def\comment#1{}

\newcommand{\be}{\begin{equation}}\newcommand{\ee}{\end{equation}}
\newcommand{\bea}{\begin{eqnarray}}\newcommand{\eea}{\end{eqnarray}}
\newcommand{\beaa}{\begin{eqnarray}}\newcommand{\eeaa}{\end{eqnarray}}
\newcommand{\ba}{\begin{array}}\newcommand{\ea}{\end{array}}
\newcommand{\bit}{\begin{itemize}}\newcommand{\eit}{\end{itemize}}
\newcommand{\ben}{\begin{enumerate}}\newcommand{\een}{\end{enumerate}}

 \newcommand{\sfrac}[2]{\raisebox{0.095ex}{\scriptsize${\frac{#1}{#2}}$}}
 
 \newcommand{\sbf}[1]{\mbox{\scriptsize\bf{#1}}}

\def\lfrac#1#2{#1/#2}

									\def\be{\begin{equation}}
\def\ee{\end{equation}}				\def\bea{\begin{eqnarray}}				\def\eea{\end{eqnarray}}
\def\bear{\begin{array}}				\def\eear{\end{array}}

												\def\x5{x^{5}}

\begin{document}

\title{Hubbard-Stratonovich Transformation:\\ Successes, Failure,
  and Cure
}

\author{Hagen Kleinert}
\email{h.k@fu-berlin.de}


\affiliation{Institut f{\"u}r Theoretische Physik, Freie Universit\"at Berlin, 14195 Berlin, Germany}
\affiliation{ICRANeT Piazzale della Repubblica, 10 -65122, Pescara, Italy}


\vspace{2mm}

\begin{abstract}
 We recall the successes of the 
Hubbard-Stratonovich Transformation (HST) of many-body
theory, point out its failure
to cope with competing 
channels of collective phenomena and show
how to overcome this by 
Variational Perturbation Theory.
That yields exponentially fast converging 
results, thanks to the help  of a variety of
  {\it collective classical fields},  rather than a fluctuating  {\it
collective
  quantum field\/} as suggested by the HST.
 
\end{abstract}

\pacs{98.80.Cq, 98.80. Hw, 04.20.Jb, 04.50+h}

\maketitle


{\bf 1}. The Hubbard-Stratonovich transformation (HST) has a well-established place in
many-body theory \cite{MBT} and elementary particle physics \cite{EPP}.
It has led to a good understanding of important collective
physical phenomena  such as superconductivity, superfluidity of He${}^
3$,  plasma and other charge-density waves, pion physics and 
chiral symmetry breaking in quark theories \cite{hqt}, etc.
It has put heuristic calculations such as the
Gorkov's  derivation \cite{GOR}  of the 
Ginzburg-Landau equations \cite{gl}
on a solid theoretical ground \cite{cqf}.
In addition, it is in spirit close \cite{CL}
to the famous density functional
theory
 \cite{DFT} via the celebrated Hohenberg-Kohn and Kohn-Sham theorems  \cite{HKT}.

The transformation is  cherished by
theoreticians
since it allows them to
re-express
a four-particle interaction {\it exactly\/}
in terms of a collective 
field variable whose fluctations 
can in principle be described by higher loop diagrams.
The only bitter pill is 
that any approximate treatment 
of a many-body system can describe interesting 
physics only 
if calcuations may be restricted 
to a few low-order diagrams.
This is precisely the point where 
the HST fails.

Trouble arises in all those many-body systems
in which different collective effects 
 compete with similar strenghts.
Historically, an important example is
the fermionic superfluid He${}^3$.
While BCS superconductivity was 
described easily via the  HST
by transforming the 
four-electron interaction 
to a 
field theory of Cooper pairs,
this approach did 
initially not succeed
in a liquid of He${}^3$ atoms.
Due to the strongly repulsive core of an atom,
the forces in the  attractive $p$-wave are not sufficient to bind the Cooper pairs.
Only after taking the help
of another collective field
that arises 
in the competing 
paramagnon channel  
into account, could the formation of weakly bound Cooper pairs 
be explained \cite{He3}.

It is the purpose of this note to point out how to circumvent 
the fatal fucussing of the HST upon a single channel 
and to show how this can be avoided
in a way that 
takes 
several competing channels into account
to each order in perturation theory.

{\bf 2}. The problem 
of channel selection of the HST was emphasized in  the context of quark theories
in \cite{hqt} and in many-body systems such as He${}^3$ in \cite{cqf}.
Let us briefly recall how it 
appears. Let $x=(t,{\bf x})$ be the time and space coordinates, 
and consider the action
$
{\cal A}\equiv {\cal A}_0+
  {\cal A}_{\rm int}$ of a nonrelativistic many-fermion
  system
\begin{eqnarray} \label{te-2.1'}
{\cal A}\!\!\!&=&\!\!\!
\int_x\psi ^*_x \left[ i \partial _t-
\xi 
   (-i \nablab  ) \right] \psi_x
-\frac{1}{2}\! \int_{x,x'}\! \psi ^*  _{\xp}\psi ^* _x
 V  _{x,x'} \psi_x\psi_{\xp},
\nonumber\\ \end{eqnarray}
where
we have written $\psi_x$ instead of $\psi(x)$, and 
$\int_{x}$ for $\int d^4x$,
to save space. The symbol
$
\xi 
   ({\bf p})\equiv
 \epsilon ({\bf p}) -\mu$ denotes the single-particle energies 
minus chemical potential.
Adding to ${\cal A}$ also a source term ${\cal A}_{\rm s}=
 \int d^4x (\psi^{*}  _x \eta_x + \cc 
)$ 
to form 
$\bar{\cal A}={\cal A}+{\cal A}_{\rm s}$,
 the grand-canonical generating
functional of all fermionic Green functions
reads 
%
  $Z [\eta, \eta^{*} ] = \int {\cal D} \psi^{*}  {\cal D}\psi
\,e^{
       i\bar {\cal A} 
}$.

The HST enters the arena by rewriting
the interaction part
with the help of an auxiliary complex field
$ \Delta_{x,x'}$ as \cite{cqf}
\\[-2em]
\begin{eqnarray}
  Z [\eta, \eta^{*} ] = \int {\cal D} \psi^{*}  {\cal D}\psi
  {\cal D}\Delta^*
  {\cal D}\Delta \,e^{
       i {\cal A} _{\rm a}  [\psi^{*} , \psi, \Delta^{*} , \Delta]
        + i{\cal A}_{\rm s} 
}
\label{4.2}\end{eqnarray}
\\[-1.em]%
       with an auxiliary action
\begin{eqnarray}
&&\!\!\!\!\!\!\!\!{\cal A}_{\rm aux}  \!     =\!
 \int_{x,x'}\bigg\{ \psi^*_x \left[ i\partial _t- \xi
        (-i\nablab)\right] \delta _{x,x'} \psi_{x'}\nonumber \\ \!\!\!\!\!\!
&&\!\!\!\!\!\!\!\!
- \sfrac{1}{2} \Delta^{*}
      _{x,x'} \psi_x \psi_ {x'}
 -\sfrac{1}{2} \psi^{*} _x \psi^{*}_{x'} \Delta  _{x,x'}
    + \sfrac{1}{2} |\Delta  _{x,x'}|^2 /V  _{x,x'}\bigg\} ,
\label{4.3}~~\end{eqnarray}
Indeed, if the field $ \Delta _{x,x'}$ is integrated out in
(\ref{4.2}), one recovers
the original generating functional. 
At the classical level, the field $\Delta _{x,x'} $
 is nothing but a convenient abbreviation for the composite {\it pair
 field} $V _{x,x'}  \psi_x \psi_{x'}$
upon extremizing the new action with respect
 to $\delta\Delta^{*} _{x,x'} $, yielding
 %
$  \lfrac{\delta {\cal A}}{\delta\Delta^{*} _{x,x'} }
  = \left(
 \Delta  _{x,x'} -
    V  _{x,x'} \psi_x \psi_{x'}
\right)/2V  _{x,x'}  \equiv 0.$
%
     Quantum mechanically, there are Gaussian
     fluctuations around this solution
     which are discussed in detail in  \cite{hqt,cqf}.

Expression (\ref{4.3}) is quadratic in the fundamental
fields $\psi_x$ and reads in functional matrix form
$
 \sfrac{1}{2} f^{*}  _x A _{x,x'} f_{x'}  $
with 
\begin{eqnarray}
   A _{x,x'} \!=\!\left(
\begin{array}{cc}
     \left[ i\partial _t-\xi(-i\nablab )\right] \delta
 _{x,x'}  &
        -\Delta _{x,x'}    \\{}
     -\Delta^{*}  _{x,x'}   &  
 \left[ i\partial _t + \xi
                 (i\nablab)\right] \delta _{x,x'} 
\end{array}
       \right) \!.
\label{4.5}\end{eqnarray}
  where
    $f_x$ denotes the fundamental field doublet 
(``Nambu spinor'') with $f^T_x
    =
\left(
\psi_x ,
\psi ^*_x
\right)$,
 and $f^\sdag\equiv f^{*T}$, as usual. Since $f^*_x$ is not independent
 of $f_x$, we can integrate out 
the Fermi fields and find
\begin{eqnarray} \label{4.6}
   Z[\eta^{*} , \eta] = \int
    {\cal D}\Delta^*
  {\cal D}\Delta \,e^{i{\cal A}[\Delta ^*,\Delta ]-\frac{1}{2}\int _{x,x'}
  j^\sdag_x[G_\Delta]_{x,x'}j_{x'}
  },
\end{eqnarray}
where  $j_x$ collects the external source $\eta_x$ and its complex conjugate,
$j^T_x
    \equiv
\left(
\eta _x  ,
\eta ^*_x
\right)$,
and the collective action reads
\begin{eqnarray}
   {\cal A}[\Delta^{*} , \Delta] = 
-\sfrac{i}{2}
            {\rm Tr} \log \left[ i {\bf G}_\Delta^{-1}
             \right]  \!+ \!\sfrac{1}{2}\!
             \int_{x.x'} \vert \Delta _{x,x'} \vert^2/
             {V _{x,x'}}.
\label{4.7}\end{eqnarray}
The $2\times2$ matrix ${\bf G}_\Delta$ denotes the propagator $iA^{-1}$
which satisfies
the functional matrix equation
\begin{eqnarray}
&& \!\!\!\!\!  \!\!\!\!\! \left(
\begin{array}{cc}
   \left[  i\partial _t - \xi (-i\nablab)\right]
          \delta _{x,x'} &  - \Delta  _{x,x'}     \\{}
    - \Delta^{*}  _{x,x'} &  
 \left[ i\partial _t +
            \xi(i \nablab)\right] _{x,x'}
\end{array}\right)
\nonumber 
%
 [{\bf G}_\Delta ]_{x',x''}
\label{4.8}\end{eqnarray}
is equal to
$ i \delta 
 _{x,x''}$.
     Writing ${\bf G}_\Delta$
     as a matrix
     $%
\left(
\begin{array}{cc}
    G_\rho &  G_\Delta   \\{}
     G_\Delta^{\sdag}  & \tilde G_\rho
\end{array} \right)
$ the mean-field equations associated with this action
are precisely
the equations used by Gorkov
 \cite{GOR}
 to study the behavior
of type II superconductors.
 %

With $Z[\eta^{*} , \eta]$
being the {\it full} partition function of the system,
the fluctuations of the collective field $\Delta _{x,x'}
$
can now be incorporated, at least in principle,
thereby yielding corrections to
these equations.

{\bf 3}. The basic weakness of the HST lies in the ambiguity 
of the decomposition of the quadratic decomposition 
(\ref{4.2}) of 
 the interaction
in 
(\ref{te-2.1'}).
For instance, there exists an
{\it alternative} elimination of
the two-body interaction 
using 
an auxiliary real field
$\varphi_x$, and writing the partition function as
\begin{eqnarray} \label{3.2}
  Z[\eta^{*} , \eta] = \int {\cal D} \psi ^{*}  {\cal D}\psi D\varphi  \exp
   \left[ i{\cal A} [\psi ^{*} ,\psi, \varphi]  + i
{\cal A}_{\rm s}
 \right],
\end{eqnarray}
rather than (\ref{4.2}),
where the action is now
\begin{eqnarray} \label{3.3}
&&\!\!\!\!\!\!\!\!\!\!\!\!
{ {\cal A} [\psi ^{*} ,\psi, \varphi]  \!  =\! \!
\int_{x,x'} \bigg\{
  \psi ^{*}_x\!
\left[
 i\partial _t \!-\! \xi
         (-i\nablab)\! -\! \varphi (x)\right]
         \delta _{x,x'}\psi _{x'} } 
\nonumber \\{}   & &~~~~~~~~~~~~~~~~~~~~~~~~~~~~~~~~~~~~~~~~~
+ \sfrac{1}{2} \varphi_x V^{-1} _{x,x'} \varphi _{x'}\bigg\}. 
\end{eqnarray}
 The  new collective quantum field $\varphi_x$ 
is directly related to the particle
 density. At the classical level, this is
obtained
from  the 
field equation 
 $ \lfrac{\delta {\cal A}}{\partial \varphi _x} = \varphi_x
   - \int dx' V_{x,x'}\psi ^{*} _{x'} \psi _{x'} = 0.$
%
For example, if $V_{x,x'}$ represents the Coulomb interaction $\delta_{t,t'}/|{\bf
  x}-{\bf x}'|$ in an electron gas, the field 
 $\varphi_x$ describes the plasmon fluctuations in the gas.

The trouble with the approach is that 
when introducing a collective quantum field
$\Delta_{x,x'}$ or $\varphi_x$,
the effects of the other is automatically included if we sum over all 
fluctuations. 
At first sight, this may appear as an advantage.
Unfortunately, this is an illusion.
Even the lowest-order fluctuation effect is 
extremely hard to calculate, already for the simplest models 
of quantum field theory such as the Gross-Neveu model, since the
propagator of the collective quantum field 
is a very complicated object.
So it is practically impossible
to recover the effects 
from the loop calculations 
with these propagators.
Thus the use of 
a collective quantum 
field theory must be abandoned
whenever 
collective effects of the different channels are important.
 
The cure of this problem comes
from the development some time ago, in the treatment
of path integrals of various quantum mechanical systems \cite{PI}
and in the calculation of critical exponents 
in $\phi^4$-field theories \cite{KS},
of
a technique  
called {\it Variational Perturbation Theory} (VPT) \cite{REMCR}.
This is democratic in 
 all competing channels 
of collective phenomena.
The important point is that it is based on 
the introduction of {\it classical collective fields}
which no longer fluctuate, and thus 
avoid double-counting 
of diagrams
of 
competing channels by quantum  fluctuations.

\def\alphan{\uparrow}
\def\betan{\downarrow}
{\bf 4}.
To be specific
let us
assume the fundamental interaction to be of the {\it local\/} form
%
\begin{eqnarray}
  {\cal A}_{\rm int}^{\rm loc} =\frac{g}2\!
     \int_x \psi _{\alpha}^{*} 
     \psi _{\beta} ^{*}  \psi _{\beta}
     \psi _{\alpha}=
  {g}\!
     \int_x \psi _{\alphan}^{*} 
     \psi _{\betan} ^{*}  \psi _{\betan}
     \psi _{\alphan}
,
\label{4.28}\end{eqnarray}
where the subscripts ${\alphan}$,
 $\betan$ indicate spin directions,
and 
we have absorbed the spacetime arguments $x$ in 
the spin subscripts, for brevity.

We now introduce auxiliary {\it classical\/}
collective fields and replace the exponential of the 
action in 
the generating functional
  $Z [\eta, \eta^{*} ] = \int {\cal D} \psi^{*}  {\cal D}\psi
\,e^{
       i\bar {\cal A} 
}$ identically by \cite{rema}
\begin{eqnarray}
\lefteqn{ \!\!
e^{i\, {g}
   \int_x \psi _{\alphan,x} ^{*}\psi _{\betan,x} ^{*}
   \psi _{\betan,x}  \psi _{\alphan,x}
     }
  =e^{-\sfrac i2\int_x f^T_x{\cal M}_{x}f_x}
 \times e^{i{\cal A}_{\rm int}^{\rm new} }}\\{}
   & & 
 \!\!\!\! \!\!\!\!
= e^{- {\sfrac i2} \!\int_x
          \left(
             \psi _{\beta}
            \Delta ^{*} _{ \beta \alpha }\psi _{\alpha}   
+ \psi _{\alpha} ^{*}
  \Delta _{\alpha \beta }
                      \psi _\beta ^{*}  
 +            \psi^ * _\beta
            \rho _{ \beta \alpha }\psi _\alpha   
+\psi _\alpha^{*}
  \rho_{\alpha \beta }
                      \psi _{\beta}   
\right)
  }\!\! \times\! e^{
i
{\cal A}_{\rm int}^{\rm new}
},\nonumber
\label{4.29a}\end{eqnarray}
with the new interaction
\begin{eqnarray}
&&
\!\!\!\!\!\!\!\!\!\!\!\!\!\!\!{\cal A}_{\rm int}^{\rm new}={\cal A}_{\rm int}^{\rm loc}
+\frac12
\int_x f^T_x{\cal M}_{x}f_x=
    \int_x \bigg[
\frac{g}2 \,\psi _{\alpha}^{*} \psi _{\beta} ^{*}
   \psi _{\beta} \psi _{\alpha}
\\
&&
+
\sfrac12 \!
          \left(
             \psi _{\beta}
            \Delta ^{*} _{ \beta \alpha }\psi _{\alpha}   
+ \psi _{\alpha} ^{*}
  \Delta _{\alpha \beta }
                      \psi _\beta ^{*}  
\right)
 +            \psi^ * _\alpha            \rho _{ \alpha\beta  }\psi
 _\beta
%
 \comment{      
      \psi _\betan
            \Delta ^{*}_{ \betan \alphan }
\psi _\alphan  + \psi _\alphan ^{*}
   \Delta _{\alphan \betan }
                      \psi _\betan ^{*}  
 +            \psi^ * _\alphan
            \rho _{ \betan\betan  }\psi _\alphan   
+\psi _\betan ^{*}
  \rho_{\alphan \alphan }
                      \psi _\betan   
}
\bigg].\nonumber 
\label{4.29}\end{eqnarray}
We now define a new 
free action 
by the quadratic form
${\cal A}_0^{\rm new}\equiv
{\cal A}_0 -\sfrac12\int _x f^T_x{\cal M}_xf_x= \sfrac{1}{2} f^{\dagger}  _x A
^{\Delta,\rho}_{x,x'} f_{x'}  $,
where   $f^T_x$ denotes the
 fundamental field doublet $f^T_x
    =
\left(\psi_\alpha ,
\psi ^*_\alpha
\right)$.
Then we rewrite 
${\cal A}_0^{\rm new}$ in the  
$2\times2$ matrix form analogous to 
(\ref{4.5}) as
${\cal A}_0^{\rm new}\equiv
{\cal A}_0 -\sfrac12\int _x f^T_x{\cal M}_xf_x= \sfrac{1}{2} f^{\dagger}  _x A ^{\Delta,\rho}_{x,x'} f_{x'}  $,
with the
 functional matrix 
$ A _{x,x'}^{\Delta,\rho}$ being now equal to
\begin{eqnarray}
 \left(\!\!
\begin{array}{cc}
     \left[ i\partial _t\!-\!\xi(-i\nablab )
\right]\!\delta_{\alpha\beta}\!+\!\rho_{\alpha\beta}
  &\!\!
        \Delta_{\alpha\beta}   \\{}\!\!
     \Delta^{*}_{\alpha\beta}
  &\!\!  \! \left[ i\partial _t \!+\! \xi 
                 (i\nablab)\right]\!\delta_{\alpha\beta}\!-\!\rho_{\alpha\beta}
\end{array}\!\!\!\right)
 \!.
\label{4.5X}\end{eqnarray}
The physical properties 
of the theory associated with the action 
$
 {\cal A}_{\rm int}^{\rm loc}+
 {\cal A}_{\rm s}$ can now be derived as follows:
first we calculate the generating functional 
of the new quadratic action ${\cal A}_0^{\rm new}$ via the functional integral
  $Z_0^{\rm new} [\eta, \eta^{*} ] = \int {\cal D} \psi^{*}  {\cal D}\psi
\,e^{
       i {\cal A} _0^{\rm new}
}$. From its derivatives we find the new free 
propagators
$G_\Delta$ and
$G_\rho$.
To  higher orders,
we expand the exponential 
$e^{i{\cal A} _{\rm int}^{\rm new}}$ in a power series 
and evaluate all expectation values 
$(i^n/n!)\langle [{\cal A} _{\rm int}^{\rm new}]^n \rangle _0 ^{\rm new}$
using Wick's theorem 
as a sum of products of the free particle propagators
$G_\Delta$ and
$G_\rho$.
The sum of all diagrams up to a certain order $g^N$
defines an effective collective action ${\cal A}^N_{\rm eff}$
as a function of 
the
collective classical fields
 $  \Delta _{\alpha \beta },\Delta ^{*} _{ \beta \alpha },
 \rho_{\alpha \beta } $,

Obviously, if the expansion is carried to infinite order,
the result must be independent of the 
auxiliary collective fields
since  they were introduced and removed in 
(\ref{4.29a})
without changing the theory.
However, 
any calculation can only be carried up to a finite order, and that will {\it depend\/}
on these fields. We therefore expect the best approximation 
to  arise from the 
extremum of the effective action \cite{PI,KS,Lecture}.

The lowest-order effective collective action 
is obtained 
from the trace of the
logarithm
of the matrix 
(\ref{4.5X}):
\begin{eqnarray}
   {\cal A} ^0_{\Delta,\rho} = -\sfrac i2 
            {\rm Tr} \log \left[ i {\bf G}_{\Delta,\rho}^{-1}
 \right]. 
\label{4.7x}\end{eqnarray}
The $2\times2$ matrix ${\bf G}_{\Delta,\rho}$ denotes the propagator
$i[A _{x,x'}^{\Delta,\rho}]^{-1}$.

\comment{    Writing ${\bf G}_\Delta$
     as a matrix
     $%
\left(
\begin{array}{cc}
    G &  F     \\{}
     F^{\sdag}  & \tilde G
\end{array} \right)
$ the mean-field equations associated with this action
are precisely
the equations used by Gorkov to study the behavior
of type II superconductors.\footnote{
 See, for example, p. 444 in Ref.~\cite{c-sc-3}}
 %
}

To first order in perturbation theory
we must calculate the expectation value 
$\langle {\cal A}_{\rm int}\rangle$ of the interaction
(\ref{4.29}). This is done with the help 
of the Wick contractions
in the three channels, Hartree, Fock, and Bogoliubov:
\begin{eqnarray}
\!\!\!\!\!\!\!\langle \psi _\alphan ^{*}  \psi _\betan ^{*}
   \psi _\betan   \psi _\alphan \rangle
\!&=&\!
\langle \psi _\alphan ^{*}  \psi _\alphan\rangle
 \langle  \psi^* _\betan   \psi_ \betan  \rangle
-\langle \psi _\alphan ^{*}  \psi _\betan\rangle
 \langle  \psi^* _\betan   \psi _\alphan  \rangle
\nonumber\\&+&
\langle \psi _\alphan ^{*} \psi _\betan ^{*}\rangle
 \langle  \psi _\betan   \psi _\alphan  \rangle.
\label{@}\end{eqnarray}
For this purpose we now introduce 
the expectation values
\begin{eqnarray}&&\!\!\!\!\!\!\!\!
\tilde\Delta^* _{\alpha \beta }\equiv g\langle \psi _\alpha ^{*}  \psi _\beta ^{*}\rangle
,~~~~\tilde
\Delta _{\beta\alpha  }\equiv g
 \langle  \psi _\beta   \psi _\alpha \rangle=[\Delta^* _{\alpha \beta }]^*
,\\ &&~\!\!\!\!
\tilde\rho_{\alpha \beta }\equiv g\langle \psi _\alpha ^{*}  \psi _\beta \rangle
,~~~~~
\tilde\rho_{\alpha\beta }^\dagger=[\tilde\rho_{\beta\alpha }]^*
,
\label{@FEXPv}\end{eqnarray}
and rewrite $\langle {\cal A}_{\rm int}\rangle$ as
\begin{eqnarray}\!\!\!\!\!
\!\!\!\!\langle {\cal A}_{\rm int}\rangle\!\!\!&=&\!\!
(\lfrac1{g}) \int_x (\tilde
 \Delta_{\betan\alphan} ^*
\tilde
 \Delta_{\betan\alphan} 
-\tilde\rho  _{\alphan\betan}\tilde \rho  _{\betan\alphan}
+\tilde\rho  _{\alphan\alphan}\tilde\rho  _{\betan\betan})
\nonumber \\
\!\!\!\!&-&\!\!(\lfrac1{2g})\int_x(\tilde\Delta_{\beta\alpha}\Delta^* _{\beta\alpha}
+\tilde\Delta^*_{\alpha\beta}\Delta _{\beta\alpha}
+2
\tilde \rho_{\alpha\beta} \rho_{\alpha\beta}
).
\label{@EXPE}\end{eqnarray}
Due to the locality of $\tilde \Delta_{\alpha\beta}$ the diagonal
matrix elements vanish and $\tilde
\Delta_{\alpha\beta}=c_{\alpha\beta}\tilde \Delta$, where
$c_{\alpha\beta}$
is $i$ times 
the Pauli matrix $\sigma^2_{\alpha\beta}$.
In the absence of a magnetic field, the expectation values $\tilde
\rho_{\alpha\beta}$
may have certain symmetries:
\\[-2em]
\begin{eqnarray}
~\tilde
\rho _{\alphan\alphan  }\equiv
\tilde
\rho =
\rho _{\betan\betan  },~~~~~~~~~
\tilde
\rho _{\alphan\betan  }
=
\rho _{\betan\alphan  }
\equiv0,
\label{@ASS}\end{eqnarray}
\\[-2em]so that (\ref{@EXPE})
simplifies to 
\begin{eqnarray}\!\!\!
\!\!\!\langle {\cal A}_{\rm int}\rangle\!=\!
(\lfrac1g)\! \!\int_x\left[
(|\tilde
 \Delta|^2\! 
+\!\tilde\rho ^ 2)
 -(\tilde\Delta\Delta^*
+\tilde\Delta^*\Delta
+2
\tilde \rho \rho
)\right]\!.
\label{@EXPE1}\end{eqnarray}
The total first-order collective classical action 
${\cal A} ^1_{\Delta,\rho} $ is given by the sum
${\cal A} ^1_{\Delta,\rho}\!\!
=\!
{\cal A} ^0_{\Delta,\rho}
\!+\!
\langle {\cal A}_{\rm int}\rangle.$

Now we observe that
the functional derivatives 
of the zeroth-order action ${\cal A} ^0_{\Delta,\rho}$
are the free-field propagators $G_\Delta$, and $G_\rho$
\begin{eqnarray}
\frac{\delta}{\delta \Delta_{\alpha\beta}}
{\cal A} ^0_{\Delta,\rho}=[G_\Delta]_{\alpha\beta}\,, ~~~
\frac{\delta}{\delta \rho_{\alpha\beta}}
{\cal A} ^0_{\Delta,\rho}=[G_\rho]_{\alpha\beta}\,.
\label{@exep2}\end{eqnarray}
Then we can extremize 
${\cal A} ^1_{\Delta,\rho} $
with respect to 
$\Delta$ and $\rho$, and find
that, to this order,
the field expectation values
(\ref{@FEXPv})
are given by the free-field propagators 
(\ref{@exep2})
at equal arguments:
\begin{eqnarray}
\tilde \Delta_x=g[G_\Delta]_{x,x} ,~~~~
\tilde\rho_x=
g[G_\rho]_{x,x}.
\label{@}\end{eqnarray}
\comment{
At zero temperature, these
are qual to 
\begin{eqnarray}
G_\Delta=\tilde \Delta=\frac{1}{V}\sum_
{\sbf p}\frac{\Delta}{E_{\sbf p}},~~~~
G_\rho=\tilde\rh=-\frac{1}{V}\sum_
{\sbf p} \frac{\xi_{\sbf p }-\mu+\rho) \Delta}{E_{\sbf p }}
\label{@}\end{eqnarray}
}
Thus we see that at the extremum, the action  
$
{\cal A} ^1_{\Delta,\rho}
$
is the same as the
extremal action 
\begin{eqnarray}
  {\cal A}_1[\Delta,\rho]=
  {\cal A}_0[\Delta,\rho]
- \frac1g \,\int_x
(|
 \Delta|^2 
+\rho ^ 2).
\label{@ExTRAc1}\end{eqnarray}
Note  how the 
theory differs, at this level, from the collective quantum field theory 
derived via the HST.
If we assume that $\rho$ vanishes identically,
the extremum of the one-loop action 
$ {\cal A}_1[\Delta,\rho]$ gives the same result  
as of the 
mean-field collective quantum field action  
(\ref{4.7}), which reads
for the 
present $\delta$-function attraction
$  {\cal A}_1[\Delta]=
  {\cal A}_0[\Delta]-
 \frac1g\int_x
|\Delta|^2 .$
 On the other hand,  if we extremize the action 
${\cal A} ^1_{\Delta,\rho}$
at $\Delta=0$, we find the extremum 
from the expression
$  {\cal A}_1[\Delta,\rho]=
  {\cal A}_0[\Delta,\rho]
- \frac1g \int_x
\rho ^ 2.
$
\comment{
This differs from the expression 
(\ref{3.8})
 for the $\delta$-function interaction 
$V$
derived
from the
HST by a factor
2.
}
The 
extremum of
the
first-order
collective classical action  (\ref{@ExTRAc1}) agrees 
with the good-old
 Hartree-Fock-Bogolioubov theory.

The essential difference between 
this and the new approach arises in two ways:\\
 \begin{itemize}
\vspace{-1.em}
\item
First when it is carried to higher orders.
In the collective quantum field theory based on the HST
the higher-order diagrams 
must be calculated with the help of the propagators of the collective 
field such as $\langle  \Delta_x \Delta_{x'}\rangle$. These are extremely 
complicated functions. For this reason, any loop diagram
formed with them is practically impossible to integrate.
In contrast to that, 
the higher-order diagrams 
in the present theory need to be calulated 
using only ordinary particle propagators
$G_\Delta$ and $G_\rho$ of Eq.~(\ref{@exep2})
and the interaction (\ref{4.29}).
Even that becomes, of course, tedious
for  higher orders in $g$.
At least, there is a simple rule 
to find the contributions
of the quadratic
 terms
$\sfrac12\int_x f^T_x{\cal M}_{x}f_x$ in  (\ref{4.29a}),
given the diagrams {\it without\/}
these terms.
One calculates
the diagrams  from 
only the four-particle interaction, and collects the contributions 
up to order $g^N$ 
in an effective action  
$\tilde{\cal A}_N[\Delta,\rho]$.
Then one replaces
$\tilde{\cal A}_N[\Delta,\rho]$ by $\tilde{\cal A}_N[\Delta-\epsilon
g\Delta,\rho-\epsilon g\rho ]$ and re-expands everything
in powers of $g$ up to the order
$g^N$, forming a new series 
 $\sum_{i=0}^ Ng ^ i \tilde{\cal A}_i[\Delta,\rho]$. 
Finally one sets $\epsilon$ equal to $1/g$ \cite{AR}
and obtains 
the desired collective classical action 
${\cal A}_N[\Delta,\rho]$ as an expansion 
extending (\ref{@ExTRAc1}):
\begin{eqnarray}~~~~~~
  {\cal A}_N[\Delta,\rho]=
\sum_{i=0}^ N\tilde{\cal A}_i[\Delta,\rho]
- (\lfrac1g)\!\int_x
(|
 \Delta|^2 
+\rho ^ 2)
\comment{
+ \frac1g\!\int_x \left[
(|\tilde
 \Delta|^2 
+\tilde\rho ^ 2)
- (\tilde\Delta\Delta^*
+\tilde\Delta^*\Delta
+2
\tilde \rho \rho)\right]
}
.
\label{@ExTRAcN}\end{eqnarray}
Note that this action must  merely be  extremized.
There are no more quantum fluctuations 
in the {\it classical collective fields} $\Delta, \rho$.
Thus, at the extremum, 
the action 
(\ref{@ExTRAcN})
is directly the grand-canonical potential.
\vspace{-.5em}
\item
The second essential difference with respect to the HST
approach 
 is that it is now possible to study a rich variety 
of possible competing collective fields
without the danger of double-counting Feynman 
diagrams. One simply generalizes the matrix ${\cal M}_{x}$
subtracted from 
${\cal A}_{\rm int}$
and added to
${\cal A}_{\rm int}$ in 
 (\ref{4.29a})
in different ways.
For instance, we may subtract 
and add a 
vector field
$\psi^\dagger \sigma^a \psi S^a$ 
 containing the Pauli matrices $\sigma^a$
and study paramagnon fluctuations, thus generalizing 
the assumption
(\ref{@ASS}) and allowing for a spontaneous magnetization 
in the ground state.
Or one may do the same thing with a term 
$\psi^\dagger \sigma^a\nabla^i \psi A_{ia}+\cc$
in addition to the previous term,
and derive the Ginzburg-Landau theory 
of superfluid He${}^3$ as in \cite{cqf}.
 \end{itemize}

An important property of the proposed procedure is that 
it yields good results in the limit of infinitely strong coupling.
It was precisely this property which led 
to the successful calculation 
of critical exponents 
of all 
$\phi^4$ theories in the textbook \cite{KS}
since critical phenomena
arise in the limit in which the unrenormalized coupling constant 
goes to infinity \cite{STR}.
This is in contrast to another possibility, in principle, of carrying
the variational approach to higher order
via the so-called {\it higher effective actions} \cite{HEA}.
There one extremizes
the Legendre transforms of the generating 
functionals of bilocal correlation functions,
which sums up all 
two-particle irreducible diagrams. 
That does  {\it not\/}
give physically meaningful results 
\cite{NOT}
in the strong-coupling limit, even
for simple quantum-mechanical models.

{\bf 6}. 
The mother of this approach, Variational Perturbation Theory
 \cite{PI}, 
is a systematic extension of a variational method
developed some years  ago by Feynman and the author \cite{fkl}.
It
converts {\it divergent} perturbation
expansions of
quantum mechanical systems into exponentially
fast {\it converging} expansions
for all coupling strength \cite{Lecture}.
What we have shown here is that this powerful theory 
can easily be transferred to
many-body theory,
if we identfy a variety of
relevant 
{\it collective   classical fields},  rather than a fluctuating collective {\it
  quantum field\/}  suggested by the HST. 
This allows us
to go
systematically beyond the 
standard Hartree-Fock-Bogoliubov approximation.

~\\
Acknowledgement:
I am grateful to
Flavio Nogueira, Aristieu Lima, 
and Axel Pelster
for intensive discussions.

\begin{thebibliography}{9}
\bibitem{rema}
{
%
Note 
that the hermitian adjoint $\Delta ^{*} _{\alphan \betan }
$ comprises transposition in the spin indices, i.e.,
  $\Delta ^{*} _{\alphan \betan } = \left[ \Delta _{\betan \alphan }
\right] ^{*} $.
}
\end{thebibliography}
 
\end{document}